# Fermi Surface and Electron Correlation Effects of Ferromagnetic Iron

J. Schäfer[1], M. Hoinkis[1], Eli Rotenberg[2], P. Blaha[3] and R. Claessen[1]

[1]*Institut für Physik, Universität Augsburg, 86135 Augsburg, Germany*
[2]*Advanced Light Source, Lawrence Berkeley National Laboratory, Berkeley, CA 94720, USA*
[3]*Institut für Physikalische und Theoretische Chemie, Technische Universität Wien, 1060 Wien, Austria*

The electronic band structure of bulk ferromagnetic iron is explored by angle-resolved photoemission for electron correlation effects. Fermi surface cross-sections as well as band maps are contrasted with density functional calculations. The Fermi vectors and band parameters obtained from photoemission and their prediction from band theory are analyzed in detail. Generally good agreement is found for the Fermi surface. A bandwidth reduction for shallow bands of ~ 30 % is observed. Additional strong quasiparticle renormalization effects are found near the Fermi level, leading to a considerable mass enhancement. The role of electronic correlation effects and the electronic coupling to magnetic excitations is discussed in view of the experimental results.

## I. INTRODUCTION

A comprehensive understanding of the electronic properties of solids has to include interactions beyond the one-electron mean-field picture. While for free-electron-like metals predictions from a simple band structure picture can be successfully obtained, the situation is much more complex for transition metals where the quasi-local character of the d-electrons leads to pronounced correlation effects. One of the phenomena that result from the electronic interaction is magnetic ordering in the low-temperature ground state. The description of ground state properties such as crystal structure and magnetic moments has been improved much since density functional theory (DFT) was introduced. Many other material properties, however, involve *excited states*. A prominent example is the determination of the electronic band structure with angle-resolved photoemission (ARPES), where *quasiparticle* energies are detected.

A model system to study electron correlations in a d-band metal is ferromagnetic iron. It is a transition metal with electron configuration $3d^6 4s^2$ that assumes a bcc crystal structure and exhibits a rather high Curie temperature of $T_C$ = 1043 K. It belongs to the group of 3d ferromagnets formed by Fe, Co and Ni, all of which contain s-p-like as well as d-like orbitals in the valence band. The d-bands are only partially filled and account for the states near the Fermi level $E_F$. In the ferromagnetically ordered state, spin-split bands with *majority* and *minority* spin character develop. In the series Fe, Co, Ni with increasing d-filling, the minority band occupation is smallest in Fe. As a result, the magnetic moment µ is highest in Fe, with a value of µ ~ 2.1 [1].

An open question is to which extent Fe is a correlated material. Self-energy effects arising from many-body interactions can by identified in the quasiparticle band structure e.g. as a reduction of the bandwidth when referenced to DFT calculations. Furthermore, a correlation satellite near the bottom of the valence band may occur, reflecting a photoexcited state where the photohole is not screened with a d-electron as in the main spectrum [2]. Experimentally, a correlation satellite in Fe at ~ 3 eV cannot easily be distinguished from the density of states and has only recently been established [3]. Concerning bandwidth and electron mass renormalization effects, a rather limited body of data from photoemission is available [4,5,6,7]. In a comparison of photoemission data with band structure calculations [7] it was concluded that correlation effects do not play a major role. Surface state contributions have been sorted out more recently [8,9,10]. The experimental resolution in these studies was rather low compared to today's standards, and many questions had to remain unresolved. Therefore, high resolution ARPES data are needed for an identification of correlation effects.

There are substantial indications that in iron electron correlation effects are strong and pronounced. In de Haas- van Alphen experiments [11], drastic mass renormalization has been derived for states near the Fermi level. The quasiparticle masses here are enhanced by up to a factor of three. Another study concentrating on surface states has identified quasiparticle renormalization on the spin wave energy scale by means of photoemission [12], and a characteristic spectroscopic signature in the self-energy was obtained. Very recently, a high-pressure phase of iron has been described which is superconducting [13]. It has been suggested that spin fluctuations are responsible for the condensation of superconducting charge carriers [14].

In this paper, we present an extensive account of high-resolution ARPES data on the (110) face of ferromagnetic iron. Using a tunable third generation light source, k-space locations have been determined with great accuracy, providing the first set of ARPES Fermi surfaces of iron. The data are contrasted with DFT cal culations. The comparison highlights the additional contribution of electron correlations to the excited *quasiparticle* state. Considerable reduction of the occupied bandwidth is observed at various locations of the band structure. Moreover, strong quasiparticle mass renormalization is observed close to the Fermi level. From the data, a k-space map with quantitative renormalization factors is obtained. Effects such as electron-magnon coupling and the available phase space for interactions is addressed.

The remainder of the paper is organized as follows: Section II contains details of both the ARPES experiment as well as the computational DFT procedure. In Section III, a detailed analysis of the Fermi surface sheets obtained from both theory and ARPES is given. Section IV focuses on the electron band dispersion along principal directions. In the discussion presented in Section V, particular attention is given to deviations between DFT and ARPES, and the role of electron correlation effects is discussed.

## II. EXPERIMENT AND CALCULATION

### 1. Experiment

Since the use of Fe crystals is disadvantageous due to tenacious impurities, Fe(110) films of very high purity were grown by electron-beam evaporation onto a W(110) substrate at a base pressure of $7 \times 10^{-11}$ mbar. The deposited film thickness was monitored with a quartz crystal microbalance *in situ*. The thickness was typically larger than 100 ML. Following the film growth, the crystal structure of the bcc Fe film was annealed at 500º C, and its quality confirmed by low energy electron diffraction. Iron is known to grow rather defect-free on W(110) despite a lattice mismatch of ~ 9.4 %. The strain is accommodated in a regular dislocation network, which occurs approximately in the first five layers, beyond which the lattice constant is fully relaxed and unperturbed [15,16].

ARPES was performed at T = 25 K at beamline 7.0.1 of the Advanced Light Source in Berkeley. The electron analyzer, type Scienta SES-100, was operated at a momentum resolution of ~0.012 Å$^{-1}$ and a total energy resolution of ~35 meV. The photon energy used for this study was of the order of 100-140 eV, resulting in a kinetic energy of the excited electrons which is far larger than the lattice potential. This justifies the assumption of free electron final states and the use of an inner potential to obtain the perpendicular momentum $k_\perp$ inside the solid.

### 2. Calculation

In the past, many different calculational approaches have been applied to obtain the band structure of ferromagnetic iron [17,18,19,20,21,22]. A density functional calculation in the local spin density approximation (LSDA) using the spin-polarized exchange-correlation of

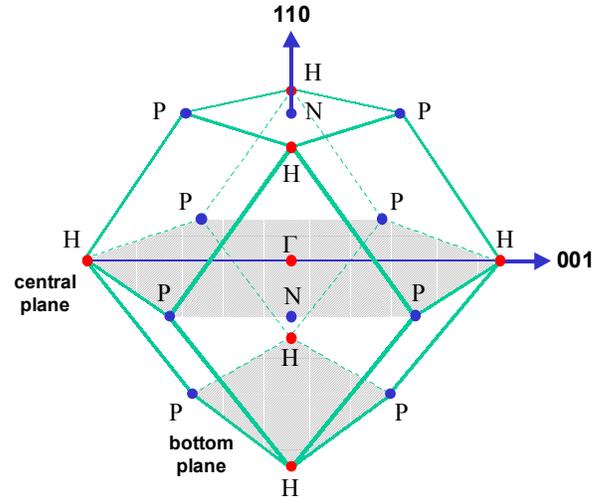

FIG. 1. Brillouin zone of bcc Fe. The central and the bottom plane parallel to the (110) surface of the crystal are accessed by the photoemission measurements.

von Barth and Hedin. was performed by Callaway and Wang in 1977 [18], and has become the "seminal" paper to which many of the following theoretical and experimental data have been compared up to now.

Here, the electronic structure of ferromagnetic iron has been calculated with a considerably more elaborate DFT approach employing the WIEN2k computer code [23]. These calculations utilize the full potential "augmented plane wave + local orbitals" (APW+LO) method, include scalar-relativistic effects and use the generalized gradient approximation (GGA) [24] to describe exchange and correlation. The GGA treatment reproduces some properties better than previous LSDA calculations. For bcc iron, the lattice constant obtained from energy optimization is too small for LSDA, while GGA yields almost the experimentally observed value [25]. In addition, LSDA wrongly predicts Fe to be more stable in a nonmagnetic fcc structure than in the experimentally observed ferromagnetic bcc one [26]. The DFT GGA output has been used to generate both Fermi surface cross sections as well as electronic band dispersions along various principal directions.

## III. FERMI SURFACES

### 1. DFT Fermi Surfaces

The Brillouin zone (BZ) of bcc Fe is a regular decahedron where each face is diamond-shaped, see Fig. 1. The corners of these faces, and hence the corners of the BZ, are formed by six H points and by eight P points. The general features of the Fermi surface (FS) of bcc Fe have been reasonably well established through the early LSDA calculation of Callaway and Wang [18]. The minority FS consists of an electron surface at Γ and six hole surfaces at the H points. The majority FS consists of a large electron surface at Γ and tubular sheets connecting the H points.

In the following we present results from our DFT calculation. FS cross sections were obtained for the (110) and (001)-planes. In Fig. 2 schematic representations of the *minority* and *majority* FS sheets for the (110) plane, parallel to the crystal face used in the experiment, are shown. By symmetry of the BZ, the same cut is also obtained for the $(1\bar{1}0)$ plane perpendicular to the crystal surface. We adopt the labeling of the FS sheets that has been used in the literature consistently since the work of Callaway and Wang [18]. The majority states form sheet *I* to *IV*, and the minority states form sheet *V* to *VIII*, as indicated in the figure.

The *minority* FS in Fig. 2(a) shows the cross section through an octahedrally shaped sheet (*VI*) around the Γ point in the center. Surrounding the H points, contours of another octahedron (*V*) are seen. These are significantly larger than the minority FS sheet in the zone center. An important detail are the two small spheres (*VII*) seen at the tips of the sheets *V* at H. The size of these spherical FS sheets depends on the exchange splitting and is rather sensitive to the computational method used, as is discussed in more detail below.

The much larger *majority* FS in Fig. 2(b) exhibits as key features a large electron octahedron (*I*) at Γ. Tubular hole FS sheets (*II*) connecting the H points are seen at the left and right side of this BZ cross section. Very near to the H point two more bands form rather small FS sheets (*III* and *IV*).

Another FS cross section in direction perpendicular to the (110) face used in the experiment is provided by the (001)-plane. The results of our DFT calculation are displayed in Fig. 3. In this cut, the BZ cross section is a square. The *minority* FS in Fig. 3(a) yields a square contour for the electron surface *VI* centered at Γ. In this view, the minute spherical FS *VII* are seen fourfold, in close vicinity to the four large octahedra *V* that surround the H points at the corners of the BZ boundary.

The *majority* FS in this (001)-cross section in Fig. 3(b) shows the rather considerable extent of the majority electron surface *I* in the center of the BZ. It almost makes contact with the tubular FS sheets *II* that connect the four H points. Again, miniscule FS sheets (*III, IV*) at the H points can be seen, resulting from additional conduction bands. We note already here that the orbital character of all states at the Fermi level is dominantly of d-character, with the only exception of the FS sheets at the N point that have some p-character. This will emerge from the band structure calculations presented below in Section IV.

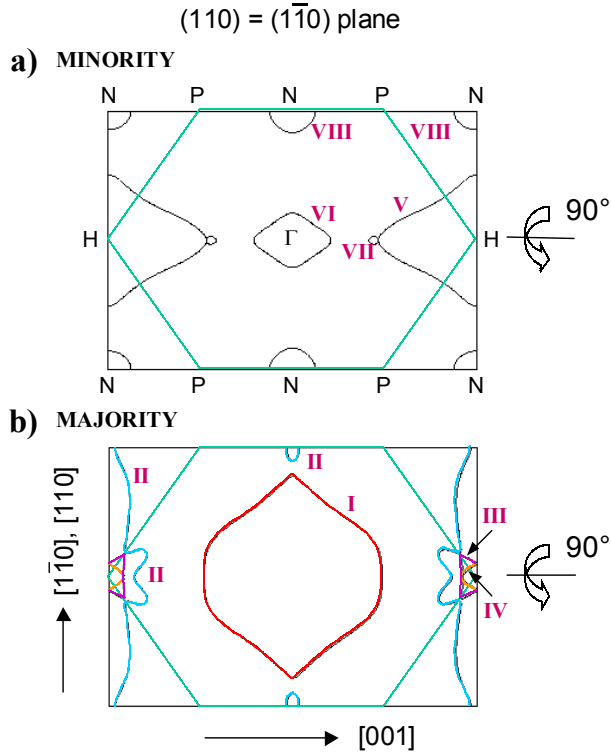

FIG. 2. DFT Fermi surface cross section in the (110) central plane. a) Minority FS sheets (*V-VIII*), b) majority FS sheets (*I-IV*). The FS cross section is the same when rotated by 90°.

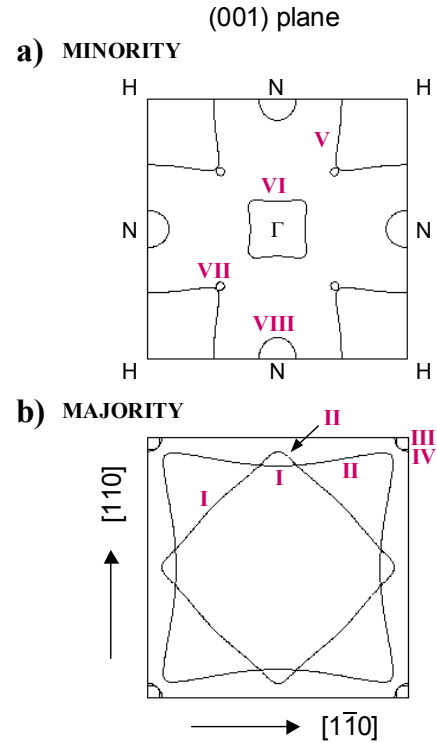

FIG. 3. DFT Fermi surface cross section in the (001) plane. a) Minority sheets, b) majority sheets. The crystal surface is in vertical direction.

## 2. ARPES Fermi Surfaces

ARPES measurements performed in order to obtain FS cross sections and band maps imply a choice of perpendicular momentum $k_\perp$ by means of selecting the photon energy. The momentum $k_\perp$ inside the solid is not conserved when the electron is transmitted across the potential barrier into the vacuum. However, in using high photon energies, one may assume free-electron-like final states, and the potential step can be modeled by a constant inner potential $V_0$. For a kinetic energy $E_{kin}$ the total momentum $k_0$ inside the crystal is therefore given by

$$\left|\overrightarrow{k_0}\right| = \frac{\sqrt{2m}}{\hbar}\sqrt{E_{kin} + V_0} \ . \quad (1)$$

where $E_{kin} = h\nu - E_B - \phi_A$. The binding energy is defined as $E_B = 0$ at the Fermi level, and $\phi_A$ is the work function of the analyzer. An ARPES angular sweep at fixed binding energy therefore follows a spherical path in k-space. The use of high photon energies (100 – 140 eV) ensures that the curvature of this sphere is kept reasonably small. Concerning the individual components of the electron momentum, the parallel momentum $k_\parallel$ does not suffer from a potential step and is given by

$$k_\parallel = \frac{\sqrt{2m}}{\hbar}\sqrt{E_{kin}} \cdot \sin\vartheta \quad (2)$$

where $\vartheta$ is the angle against the sample normal. The perpendicular momentum $k_\perp$ carries the remainder of the kinetic energy and is affected by the potential step:

$$k_\perp = \frac{\sqrt{2m}}{\hbar}\sqrt{E_{kin}\cos^2\vartheta + V_0} \ . \quad (3)$$

Technically, an automated sweep of $k_\parallel$ is performed with the detector energy window set to the Fermi level. In addition, the data acquisition technique was extended such that the full energy spectrum for each k-point was recorded. This is made possible by an imaging detector array that simultaneously records a window of energy and angles. In this manner, the principal directions of the BZ can be determined reliably from the FS data, and then the band maps of interest can be extracted as presented further below.

*a) Photon energy scan.* In order to determine the location of the symmetry points correctly, a photon energy scan was performed, ranging from 75 - 210 eV. This is important to ensure that the plane containing the Γ point is intersected correctly. For orientation, a calculated *vertical* cut of the majority FS with the surface direction (110) upward is shown in Fig. 4(a). The weakly curved paths sampled by ARPES angle scans are indicated for typical photon energies used in this paper.

Data from the photon energy scan used to identify the Γ point are shown in Fig. 4(b). In order to obtain high intensity for both minority and majority FS surrounding the Γ point, the plane chosen for this scan was rotated around its vertical axis by 35.3º into the $[1\bar{1}2]$ direction pointing from Γ to the far N point. The ordinate of the photon energy scan in Fig. 4(b) is the photon energy, on a scale of $k_{h\nu} = \sqrt{2m}/\hbar \cdot \sqrt{h\nu}$. Although the inner potential $V_0$ is not known *a priori*, this scale serves as close approximation to $k_\perp$. At the Fermi level, $h\nu = E_{kin} + \phi_A$ is almost the same argument under the square root for the high photon energies used. This scale visualizes the symmetry in the BZ, the distortion being negligible in the vicinity of any given symmetry point, while it remains fully exact when read as photon energy. For normal emission, we observe the Γ point at $k_{h\nu} = 5.96$ Å$^{-1}$ equivalent to a photon energy of $h\nu = 135.3 \pm 2$ eV. This implies an inner potential of $V_0 = 14.3$ eV. Using an approximate sample work function of $\phi_S \sim 4.5$ eV, this corresponds to a muffin-tin zero of $E_0 = V_0 - \phi_S \sim 9.8$ eV referenced to the Fermi level. This is in reasonably good agreement with the bottom of the valence bands obtained from the DFT calculation shown below.

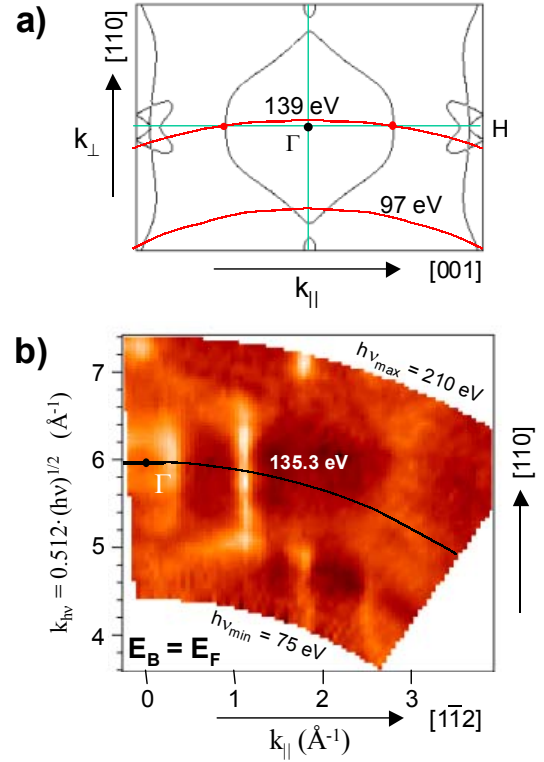

FIG. 4. a) Spherical measurement paths resulting for the photon energies utilized in this study, indicated in the DFT majority FS for an angle scan along [001]. b) Determination of the photon energy that intersects the Γ point in normal emission (angle scan along $[1\bar{1}2]$), color scale indicates states at $E_F$ in bright intensity).

*b) FS central plane.* The key Fermi surface for the (110)-oriented iron crystal face is the cut through the central plane of the BZ that includes the Γ point (plane

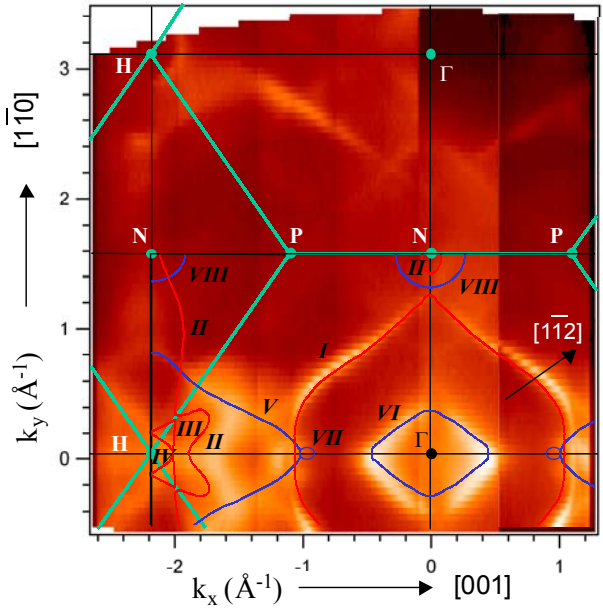

FIG. 5. ARPES Fermi surface at BZ central plane (hv = 139 eV). The experimental data are overlaid with the DFT calculation for minority and majority FS sheets. For discussion see text.

Γ-H-N), as shown in Fig. 5. The curvature of the sphere described in k-space with this photon energy vector is reasonably small because the photon energy is so high compared to conventional UV spectroscopy in the 20-40 eV range. By choosing a marginally higher photon energy of 139 eV, the weak curvature is corrected such that the central (110)-plane through Γ is intersected at $k_{||}$ = 1.0 Å$^{-1}$ away from Γ, see Fig. 4(a). This is also the average radius of the majority contour *I* surrounding Γ, and hence enables a rather exact determination of Fermi vectors in this area.

In Fig. 5 the two contours of minority and majority FS are clearly identifiable. In comparing it to the DFT FS (see Fig. 2 and overlay on the data in Fig. 5), the small diamond-shaped contour *VI* is identified as the minority contour, while the large contour *I* is the majority contour. Along the Γ-H axis, and virtually intersected by the majority contour, one finds a small circular contour, which from the calculation relates to the minute spherical contour *VII* in the vicinity of the very large minority contour *V* at H. The fact that the scan intersects these minute minority spheres is direct proof that the photon energy was chosen correctly, which otherwise would have missed this minute detail.

At the H point, we find a large diamond-shaped contour line that is the minority hole-FS *V* there. In addition, the small cauliflower-like contours predicted by the calculation for the majority contours (*II, III, IV*) in Fig. 2(b) are also identified. However, they appear somewhat closer to the large majority contour around Γ which we ascribe to the deviation of our spherical measurement area. At H we estimate the deviation from the Γ-H-N plane to be about 5% of the BZ height, therefore the scan will intersect the tubular FS sheets *II* connecting the H points at somewhat smaller $k_{||}$-values.

Concerning the minority FS sheet *VIII* and inside of it the majority FS sheet *II* at the N-point at the top end of the plot, we find that the intensity is much suppressed which may be due to matrix element effects. However, some faint contour that seems to extend the lines of the Γ majority contours can be seen. This close vicinity of the Γ majority contour (*I*) with the N minority contour (*VIII*) is consistent with the DFT calculation.

*c) FS bottom plane.* A FS cross section further away from the Γ-H-N plane is shown in Fig. 6. It is recorded with a photon energy of hv = 97 eV. This corresponds to being close to the bottom P-H-N-plane of the BZ, see Fig. 4(a). We note in passing that for both photon energies used for the FS data surface states have a vanishing cross section and thus do not show up in the data. For normal emission, the total momentum here is still ~40% away from N on the vertical Γ-N line. For large parallel momentum, however, the scan is rather close to the bottom plane, and intersects it just beyond the H point. The ARPES FS data of Fig. 6 give a particularly clear-cut view of the minority FS hole sheet *V* surrounding H. The small minority sheet *VII* is only faintly seen. The tubular FS sheets *II* are intersected such that four circles are overlaid onto the large minority sheet *V* surrounding H. These four FS sheets extend upwards to the H points in the central plane. The H point is furthermore surrounded by small majority contours *III* and *IV*, which in the data give rise to a circular structure of moderate intensity.

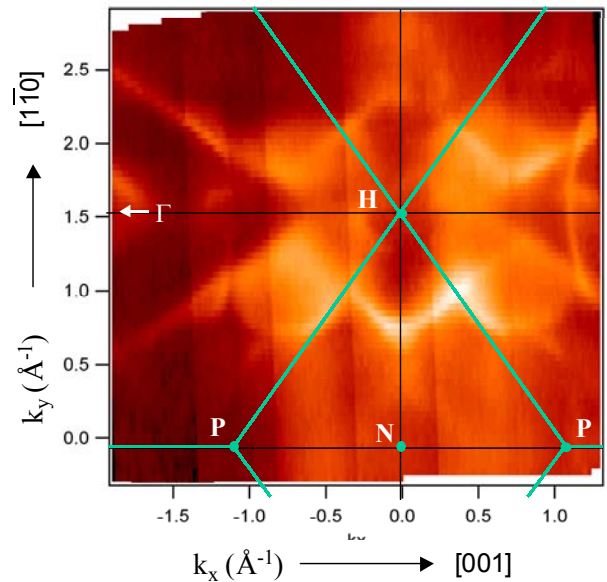

FIG. 6. ARPES Fermi surface at BZ bottom (hv = 97 eV). The diamond-shaped minority FS sheet *V* is seen with high intensity.

The lower H point also has Γ points as neighbors. At those large $k_\parallel$-values, the measurement area is close in $k_\perp$ to the far Γ point, so that its two centered FS sheets can be seen (*I* and *VI*). Combining the experimental ARPES results of Fig. 5 and Fig. 6, a rather good general agreement with the calculated FS topology emerges. A detailed analysis will be given below in conjunction with the band dispersions obtained from both theory and experiment.

### IV. BAND DISPERSIONS

**1. DFT Band dispersion**

The dispersion of the electron bands for both minority and majority electrons has been calculated with DFT. The results obtained for various principal directions are shown in Fig. 7. The valence electron configuration of Fe may be denoted as $3d^6 4s^2$. This is reflected in the band structure which shows the simultaneous presence of broad nearly-free (sp-like) and narrow quasi-local (d-like) bands. The band complex covers the energy range from -4.5 to +0.5 eV and from -3.0 to +2.5 eV for majority and minority states, respectively, and hence contributes strongly to the Fermi surface.

The exchange splitting between minority and majority band depends both on the binding energy and the k-location. At the pronounced valence band minimum at Γ with a binding energy of 8.3 eV, the exchange splitting amounts to only $U_{ex} = 0.15$ eV. In contrast, the exchange splitting in the d-band region near $E_F$ ranges from 2.2 to 2.9 eV.

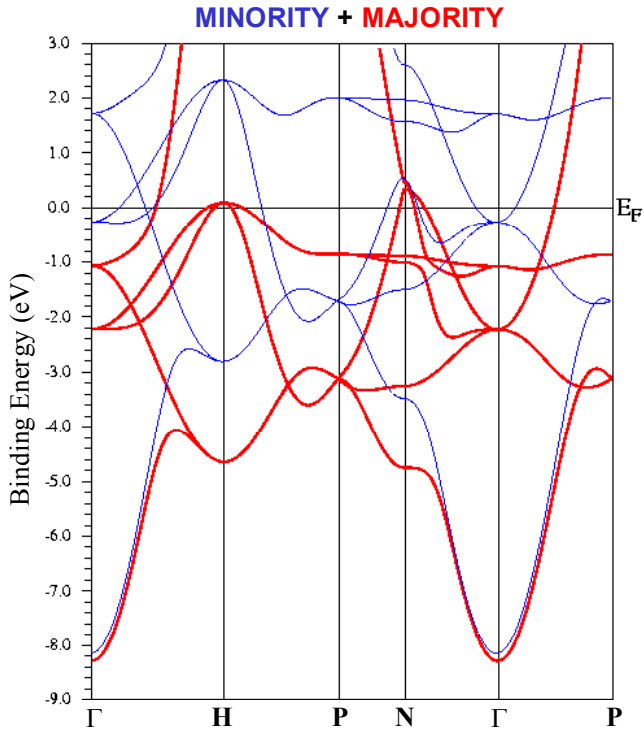

FIG. 7. DFT-GGA band structure for minority and majority spin electrons (thin and thick line, respectively). The exchange splitting is energy- and k-dependent.

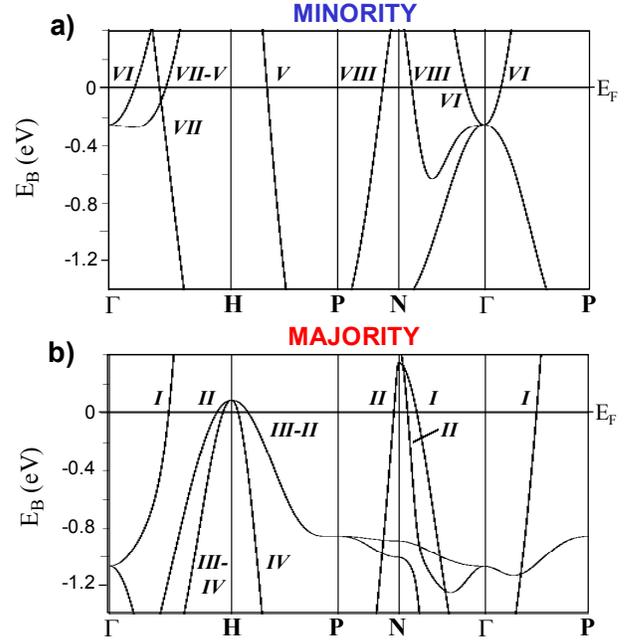

FIG. 8. DFT band structure in a close-up display near the Fermi level for a) minority and b) majority spin. Labels in roman numerals (*I-VIII*) indicate to which FS sheet the band contributes, including possible degeneracies along certain symmetry lines.

Close-up panels for minority and majority electrons are shown in Fig. 8 in an energy window of 1.4 eV below $E_F$. The data are plotted separately by spin character in order to enable determination of Fermi vectors and band slopes for comparison with experiment.

The dominant d-character of the conduction bands near $E_F$ is shown in Fig. 9(a). The energy range of the occupied d-bands is found to be ~ 3 eV for *minority* and ~ 5 eV for *majority* states. At higher binding energies, all states are of s-character. The only symmetry point with partial p-character is the N point, for both minority and majority FS sheets. Furthermore, as shown in Fig. 9(b), we find almost all d-states at $E_F$ are of $t_{2g}$-symmetry, with the only exception of $e_g$-symmetry along the Γ-H-line.

**2. ARPES Band Maps**

In the following we show band maps taken with ARPES along most principal directions. The directions shown in the band structure calculations of Fig. 7 and Fig. 8 can all be accessed in the (110)-central plane containing Γ, and have been recorded with a photon energy of $h\nu = 139$ eV, as used for the FS data.

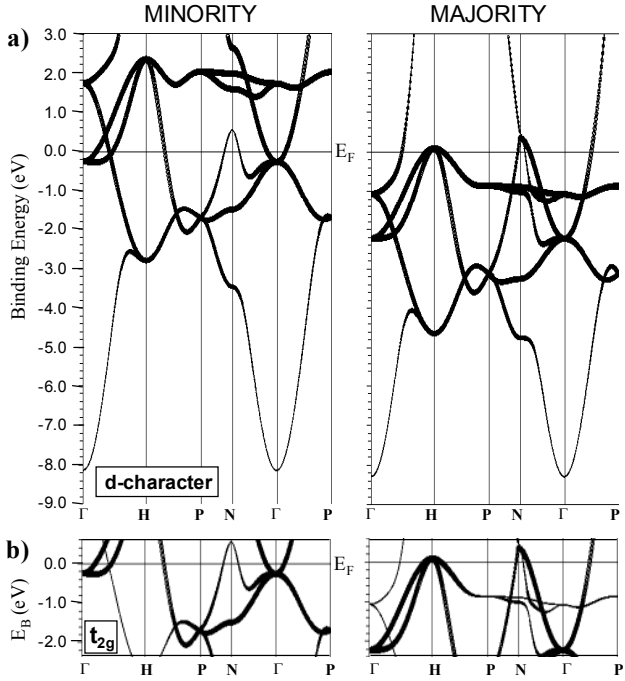

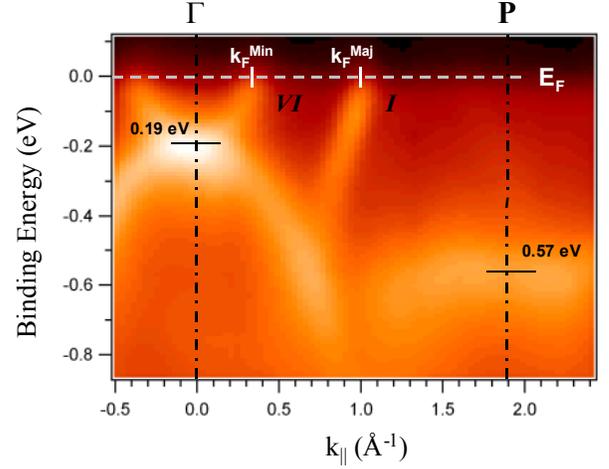

FIG. 10. ARPES band map along Γ-P (hν = 139 eV). In addition to the two Fermi vectors of sheet *I* and *VI*, two band minima at Γ and P are identified.

Fig. 9. a) Calculated d-orbital character of the Fe energy bands (indicated by circle size). This accounts for most of the weight near $E_F$. b) d-t2g orbital symmetry, typical for most of the d-bands.

*a) Band map Γ-P.* The ARPES band map along the Γ-P direction is shown in Fig. 10. This path intersects both minority and majority Fermi level crossings, at $k_F^{Min}$ = 0.32 Å$^{-1}$ and at $k_F^{Maj}$ = 0.97 Å$^{-1}$ corresponding to minority sheet *VI* and majority sheet *I*, respectively. A band minimum is seen for the minority band while for the majority band the intensity is lost at higher binding energy. In addition, a flat band is observed at the P point, with a binding energy of $E_{min}$ = 0.57 eV. It relates to the flat, d-like band seen in the band calculation that does extend to the Γ point, however, does not show up there due to a lack of intensity (even on a larger binding energy scale than shown). The photoemission cross section may be considered responsible for this behavior, yet correlation effects can furthermore lead to a considerable broadening.

*b) Band map N-Γ.* In Fig. 11 a band map scan along the Γ-N direction is plotted. In this display one can again see two Fermi level crossings corresponding to the minority sheet *VI* and majority sheet *I*, however, the intensity is generally suppressed along this direction. Yet the intensity is rather high at the Γ point, and one can easily identify the minority band at $E_{min}$ = 0.19 eV binding energy. The majority band, in contrast, looses intensity as it disperses downward, and cannot be followed to its minimum. Very close to N, two more crossings (minority sheet *VIII*, majority sheet *II*) are expected, but are not observed as the photoemission cross section turns out to be so weak here.

*c) Band map Γ-H.* ARPES data for the Γ-H direction are reproduced in Fig. 12. From the band calculation, six Fermi level crossings are expected. The pair of bands for minority FS sheet *VI* and majority FS sheet *I* that disperse upward from the Γ point are seen clearly. At $k_∥$ marginally larger than Fermi level crossing *I*, there is a crossing that we ascribe to both sheets *V* and *VII* which have a degenerate $k_F$ here. Note that in the FS data of Fig. 5 we find the spherical sheet *VII* to be larger than calculated, and penetrating sheet *I*. The sphere *VII* is intersected twice by the Γ-H line, yet photoemission intensity is obtained only for that $k_F$ closer to H.

Concerning the remaining two crossings, intensity is seen near the H point that must be ascribed to the two majority bands rising above the Fermi level there. While one of these bands is identifiable (sheet *II*), the second one (which along Γ-H is a degenerate band derived from sheets *III* and *IV*) cannot be extrapolated clearly towards $E_F$. In particular, both bands loose intensity rather rapidly for higher binding energies near ~ 0.5 eV and beyond.

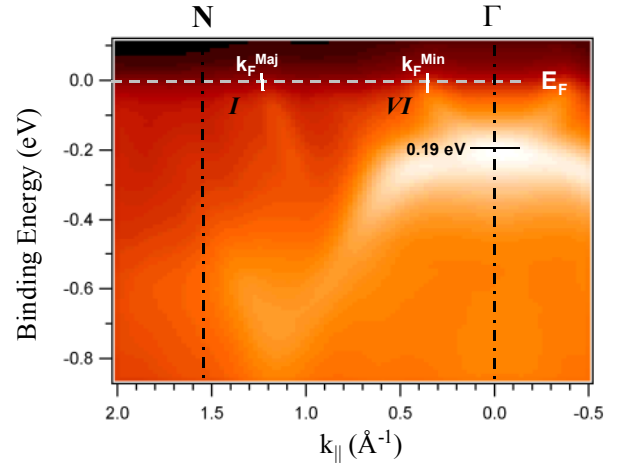

FIG. 11. ARPES band map along N-Γ (hν = 139 eV). The intensity at $E_F$ is rather suppressed for this direction. Fermi vectors for sheet *I* and *VI* are still seen.

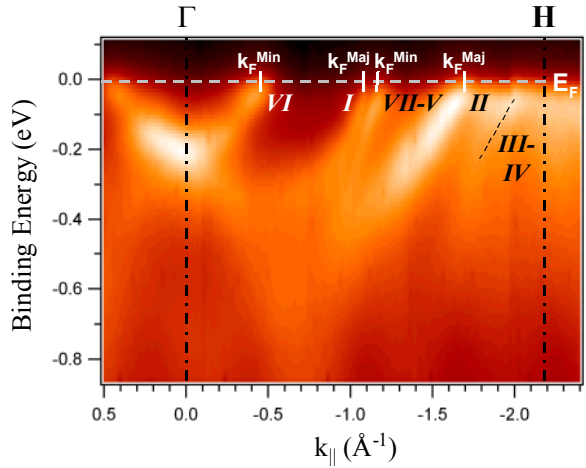

FIG. 12. ARPES band map along Γ-H (hν = 139 eV). In addition to the Fermi vector for minority FS sheet *VI*, two Fermi vectors for majority sheets *I* and *II* are observed. The splitting of the band for *I* relates to the minority FS *VII* (see also FS data Fig. 5).

*d) Band map H-P.* The band dispersion along the H-P direction is shown in Fig. 13. Well away from H towards P at $k_F = 0.68$ Å$^{-1}$ a band is seen which, from comparison with the DFT calculation, we relate to minority FS sheet *V*. Rather close to H, a band is observed which matches the shape of an inverted parabola with a maximum very close to $E_F$. It relates to the band of majority sheet *IV* forming a very small surface. Due to the increasing deviation of $k_\perp$ for large angles which begins to play a role at H, the true Fermi level crossing is likely to be slightly missed for this sheet. In-between the two intense bands, there is a faint hint of a third band. It is well explained by the degenerate band derived from majority sheets II and III. This second majority band of higher effective mass will cross the minority band *V*. In the data, the shallow band is once more observed around the P point with a minimum at 0.57 eV binding energy.

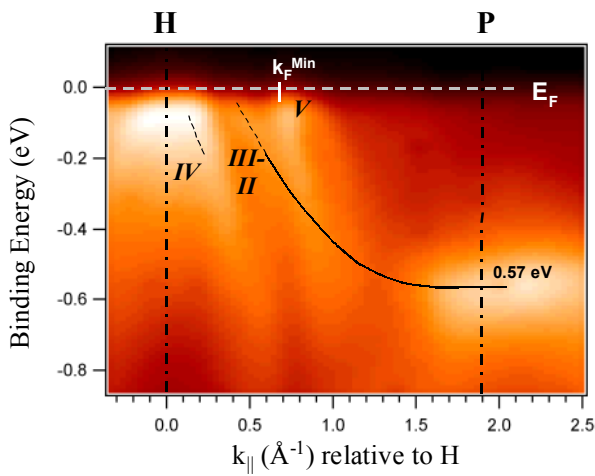

FIG. 13. ARPES band map along H-P (hν = 139 eV). The Fermi vector of FS sheet *V* is seen. Close to H, high intensity stems from two majority bands which are not expected to reach $E_F$ on the curved k-path probed by ARPES.

## V. DISCUSSION

### 1. Identification of Correlation Effects

Though DFT includes correlation and exchange contributions to the ground state (and only approximately so via LDA or GGA), the DFT energy bands representing the single-particle excitations are obtained only in a mean-field-like fashion. Thus, any deviation of the experimentally observed *quasiparticle* excitation states from the DFT bands has to be attributed to correlation effects in the electronic self-energy. In particular, the dressing of the photohole with virtual electron-hole (or other elementary) excitations will lead to a renormalization of the quasiparticle energies, thus affecting the Fermi velocity, the effective mass, or even the entire occupied bandwidth with respect to the DFT band structure. In the following we give a detailed comparison of experimental and DFT results on Fermi vectors, bandwidths, and Fermi velocities in order to identify such electronic correlation effects.

*a) Fermi vectors.* A systematic list of the Fermi vectors $k_F$ obtained from DFT and from experiment is given in Table I in the first pair of data columns. Concerning the Fermi vectors, there is overall very close agreement between theory and experiment. Allowing an error of ±0.02 Å$^{-1}$ for the ARPES readout, most values are well compatible within error bars.

Along the Γ-H line the Fermi level crossing of the large majority octahedron *I* is rather well reproduced in DFT with $k_F = 1.05$ Å$^{-1}$, compared to an ARPES value of 1.08 Å$^{-1}$. Note that the photon energy was chosen to most exactly intersect the H-Γ-N plane at this Fermi vector of sheet *I*. However, the small minority Fermi surface sheet *VII*, which is calculated to be fully inside the majority Fermi surface *I*, is found to actually overlap with the FS contour *I*. This is a second minor deviation between ARPES and DFT. The band situation here is very complicated, see Fig. 8, and the size of the FS sheets critically depends on the position of the Fermi level as well as possible hybridization effects. In particular, spin-orbit splitting not considered in the calculation may be the source of deviations. We note that in the early calculation of Callaway [18], the small FS sphere *VII* does indeed overlap with majority contour *I*, and is also slightly larger than obtained from the DFT. On the Γ-H line close to H (sheet *II*) we must expect a larger deviation between DFT and ARPES Fermi vector, as here the spherical measurement path starts to deviate noticeably from the central (110) plane. Overall, however, the variance of the calculational DFT Fermi vectors is generally rather minute when compared to ARPES. We hence conclude that despite the pronounced energy renormalization discussed below, the experimental Fermi surface is well reproduced by band theory in accordance with Luttinger's sum rule.

TABLE I. Band parameters obtained from DFT and ARPES, comprising Fermi vectors, Fermi velocities and mass renormalization ratios (experimental error margin ± 20%). The FS sheets to which the parameters apply are indicated. Some bands are not observed in ARPES ("n. obs.").
*) Note: deviating sequence of crossings in DFT vs. ARPES, prohibiting a meaningful comparison for this band.

| Direction | Spin | FS | $k_F$ DFT ($\text{Å}^{-1}$) | $k_F$ ARPES ($\text{Å}^{-1}$) | Slope $v_0$ DFT (eV/Å$^{-1}$) | Slope $v_R$ ARPES (eV/Å$^{-1}$) | Mass Renorm. Ratio $v_0/v_R$ |
|---|---|---|---|---|---|---|---|
| Γ-P | Min. | VI | 0.30 | 0.32 | 1.73 | 0.88 | 2.0 |
|     | Maj. | I  | 0.94 | 0.97 | 5.00 | 1.40 | 3.6 |
| Γ-N | Min. | VI | 0.32 | 0.36 | 1.47 | 0.80 | 1.8 |
|     |      | VIII | 1.30 | n. obs. | 3.18 | — | — |
|     | Maj. | I  | 1.21 | 1.22 | 2.37 | 1.16 | 2.0 |
|     |      | II | 1.39 | n. obs. | 5.19 | — | — |
| Γ-H | Min. | VI | 0.46 | 0.46 | 1.00 | 0.72 | 1.4 |
|     |      | VII | 0.90 | n. obs | 3.23 | — | — |
|     |      | VII-V | 1.02 | 1.18*) | 1.18 | 1.12*) | — |
|     | Maj. | I  | 1.05 | 1.08 | 3.78 | 1.12 | 3.4 |
|     |      | II | 1.90 | 1.70 | 0.72 | 0.67 | 1.1 |
|     |      | III-IV | 2.00 | n. obs. | 1.14 | — | — |
| H-P | Min. | V  | 0.66 | 0.68 | 5.38 | 1.79 | 3.0 |
|     | Maj. | IV | 0.13 | n. obs. | 1.64 | — | — |
|     |      | III-II | 0.35 | n. obs. | 0.58 | — | — |

*b) Band energies at critical points.* Band energies may also be compared between DFT and ARPES. As mentioned before, for some bands the band minimum cannot be observed because the intensity fades out for higher binding energies. The complete loss of intensity in the quasiparticle peak obtained with ARPES has also been reported for cobalt, where the suppression is particularly pronounced for binding energies higher than 2 eV [27].

For the band minima with not so high binding energies, from the close-up band structure panels of Fig. 8 we infer that shallow band minima are expected at the Γ point, at the P point, and along the Γ-N direction. Unfortunately, the Γ-N line has a very low cross section, as seen e.g. in the Fermi surface of Fig. 5, and the multiple band situation is not reflected in the data. For the Γ and P symmetry points, accurate readouts are obtained from the data with high photoemission intensity: i) For the *minority* band at Γ (FS sheet *VI*), the occupied bandwidth amounts to 0.26 eV in DFT and 0.19 eV in ARPES, corresponding to a bandwidth reduction of 27 %, i.e. a renormalization ratio of 1.4. ii) For the *majority* band at P, the occupied bandwidth amounts to 0.85 eV in DFT and 0.57 eV in ARPES, corresponding to a bandwidth reduction of 33 % equivalent to a renormalization ratio of 1.5.

We thus observe a sizable reduction of bandwidths with respect to the DFT results, attributed to an energy renormalization induced by strong electron-electron interaction in the Fe 3d conduction bands.

*c) Mass renormalization.* The Fermi velocity of the metallic bands is a further indicator of interactions experienced by the electrons that are not included in the DFT calculation. We have determined the Fermi velocities $v_F$ for both theory and experiment as listed in Table I. For the ARPES data, peak positions of the momentum distribution function have been used to determine the dispersion and its derivative as accurately as possible. The error bar of this procedure is estimated to be ± 15% for the experimental data.

Mass renormalization factors can easily be derived as the ratio of the experimental and calculated Fermi velocities, and are listed in Table I. Values ranging from 1.1 to 3.6 are observed, with a total error of the order of ± 20 %. The observed mass renormalization ratios are therefore significantly larger than unity, even admitting the experimental uncertainty. A comparison of effective electron masses from de Haas-van Alphen experiments with band structure calculations performed by Lonzarich [11] also leads to mass renormalization ratios in the range of 1.5 to 3, consistent with our ARPES data.

These numbers determined near the Fermi level are larger than the ARPES bandwidth renormalization of 1.4 – 1.5. If it were a uniform scattering mechanism, a roughly constant renormalization ratio for the whole energy range below $E_F$ would be expected. A significantly larger renormalization ratio at low binding energies, however, points at an additional interaction mechanism.

**2. Discussion of Renormalization**

*a) Systematics of renormalization.* The data show no evident systematics such that either spin type (*minority* or *majority*) would be renormalized considerably stronger than the other. The largest renormalization ratios of three and more are observed at Γ-P and Γ-H for majority carriers, while at H-P this value is assumed by the minority spins. A larger scattering rate for *minority* states has been observed in Gd for bands near the Fermi level by Valla *et al.* [28] and is explained by the higher majority density of states available for scattering into minority holes. For surface states of Fe, unoccupied states seen in inverse photoemission showed longer living majority than minority states [29]. Scattering asymmetries were also discussed by Hong *et al.* in conjunction with interaction of electrons with spin waves [30, 31]. Unlike Valla *et al.*, their argument implies a stronger scattering for holes in a *majority* band, which will be filled by minority electrons under emission of a spin wave that compensates the increase in magnetization. Our data, however, do not support preferential interactions for either spin channel.

Another obvious relationship to look for is the orbital character of the states at $E_F$. Here we refer to Fig. 9 and recall that d-$t_{2g}$ character is predominant there. As there is little change in the orbital symmetry across the FS, it cannot explain the considerable variance in the mass renormalization observed.

Rather, it appears as general trend that large mass renormalizations are observed for large Fermi surface sheets. This is visualized in Fig. 14, where those points on the FS sheets are marked where determinations of the mass renormalization have been made. The behavior may imply a complex k-dependence, yet the most obvious statement would be that more scattering channels are available for electrons on a large FS sheet. Scattering occurs predominantly with spins of opposite sign due to the larger interaction strength, as argued by Monastra *et al.* for Co [27]. One may speculate that the vicinity of FS sheets of opposite sign and much smaller occupied fraction serves as large phase space of *unoccupied* states where opposite spin electron-hole pairs can be created abundantly.

*b) Theoretical treatment of self-energy effects in Fe.* Attempts have been made to incorporate the interaction between the electrons, as observed in the ARPES spectra, in a theoretical computation of a *quasiparticle* band structure [32,33,34,35,36,37]. It is found that quasiparticles in Fe beyond 1 eV are strongly damped [32]. Inclusion of the on-site Coulomb interaction of localized electrons. [34] reproduces the correlation satellite in the spectra of Fe, Co, and Ni. It leads to an interaction energy U which for Fe is not much smaller than for Ni, in contrast to old photoemission experiments which suggested rather weak effects for Fe. Moreover, inclusion of electron correlations in Fe using LDA-DFT plus a dynamic mean-field approximation requires U to be as large as 4 eV to provide reasonable agreement with photoemission spectra [35].

In a Green function approach [36] quasiparticle renormalization factors are obtained which for the d-electrons are 1.7 in Ni and 1.4 in Fe, suggesting strong many-body interactions in both materials. In addition, a massive broadening of the density of states is observed beyond 1-2 eV binding energy. A related treatment using the GW method [37] generates a rather sizeable bandwidth reduction of the order of 25 % which is comparable to our data. These theoretical approaches are thus in support of a significant role of correlations for ferromagnetic iron.

*c) Comparison to Ni.* Bandwidth renormalization has long been known to occur in Ni [38]. The effect has lately been studied in more detail with modern high-resolution ARPES [39,40,41], and a bandwidth renormalization for the 3d bands of 27-30 % is observed. Also, a correlation satellite $\sim$ 6 eV below $E_F$ has been established for Ni from photoemission data [2]. Our current ARPES data suggest a very similar situation for Fe. We note that this finding relates to the bands close to the Fermi level, where we observe bandwidth reductions of $\sim$30 %. Deeper binding energies in Fe were explored in older work, finding a bandwidth reduction of about 10 % there [5]. For the sp-like bands of Ni, the effect there is also rather small [38]. From this comparison it emerges that the bandwidth renormalization effects in Fe are not significantly smaller than in Ni, and the similarities of these two ferromagnets are rather striking.

*d) Relevance of magnetic excitations.* While the bandwidth renormalization is thus well accounted for by interelectronic Coulomb interaction, the enormous renormalization factors up to $\sim$3 observed near the Fermi energy indicate the presence of additional scattering processes at low excitation energies. A possible candidate is the coupling between conduction electrons and spin excitations.

Recently, the formation of quasiparticles due to electron-magnon coupling has been identified in the ARPES spectrum of Fe(110) surface states by means of their particular self-energy signature [12]. Interaction of electrons with magnetic excitations in Fe films has also been observed in the spin-polarized electron energy loss spectroscopy [42,43], leading to structure at 250 meV and below.

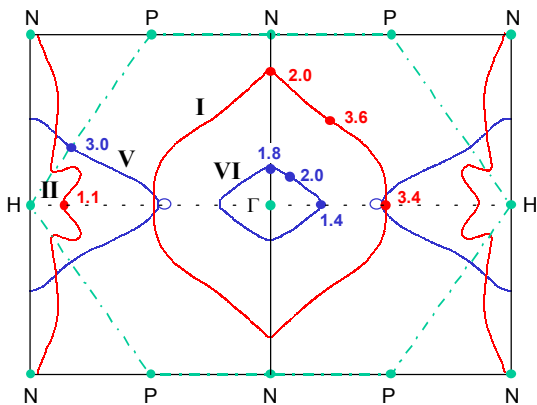

FIG. 14. Fermi surface from DFT with those sheets where mass renormalization was determined (minority: blue, majority: red). Locations and mass renormalization ratios are indicated by dots. Larger FS sheets appear to have generally larger ratios.

In fact, a pronounced coupling between electrons and spin excitations may also be the driving force for the superconducting state recently discovered in a non-magnetic high-pressure phase of iron [13]. Here, the attractive interaction between electrons needed to form Cooper pairs is likely to be mediated by such spin fluctuations [14].

## VI. CONCLUSIONS

In the pursuit of identifying correlation effects in ferromagnetic iron, ARPES has been used to establish high-accuracy Fermi surfaces for the first time, together with the quasiparticle band structure along various principal directions. Essential features of the Fermi surface can be rather accurately described with the DFT-GGA method. Many-body effects become obvious when the ARPES quasiparticle dispersions in the d-band region are analyzed. A considerable band with reduction of about 30 % compared to DFT is observed. These renormalization effects imply significant correlation effects due to electron-electron interaction. The bandwidth reduction is consistent with expectations based on various theoretical treatments for Fe. Surprisingly, much stronger mass renormalization is found very close to the Fermi level, with a drastic mass enhancement of up to a factor of three or more. In this low energy range, coupling to spin wave excitations seems to play an important role.

From these results we conclude - in contrast to earlier ARPES studies - that electronic correlations in Fe are just as strong as in the other two ferromagnetic *3d* metals Ni and Co. As an outlook and challenge to solid-state theory, it would be highly desirable to have a microscopic description of the quasiparticle band structure of Fe which is capable of reproducing the mass renormalization effects especially at the Fermi level.

## ACKNOWLEDGMENT

The authors are grateful to A. Bostwick and H. Koh for technical support. This work was supported by the DFG (grant CL 124/3-3 and SFB 484), and by the BaCaTeC program.